\documentclass[aps,showpacs,preprintnumbers,amsmath,amssymb,preprint]{revtex4}
\usepackage{graphicx}
\usepackage{dcolumn}
\usepackage{multirow}
\usepackage{slashbox}
\usepackage{subfigure}
\usepackage{bm}

\begin{document}

\title{The Neutron Energy Spectrum Study from the Phase II Solid Methane Moderator at the  LENS Neutron Source}

\author{Yunchang Shin}%
 \email{yunshin@indiana.edu}
 \homepage{http://www.iucf.indiana.edu}
 \author{Christopher M. Lavelle}
\author{W. Mike Snow}
\author{David V. Baxter}
\author{Xin Tong}
\author{Haiyang Yan}
\author{Mark Leuschner}
\affiliation{Department of Physics, Indiana University/IUCF, 2401
Milo B. Sampson Lane, Bloomington, IN 47408, USA}

\date{\today}

\begin{abstract}
Neutron energy spectrum measurements from a  solid methane moderator were performed at the  Low Energy Neutron Source (LENS) at Indiana University Cyclotron Facility (IUCF) to verify our neutron scattering model of solid methane\cite{yun2007-1}. The time-of-flight method was used to measure the energy spectrum of the moderator  in the  energy range of 0.1$meV\sim$ 1$eV$.  Neutrons were counted with a  high efficiency $^{3}{He}$ detector. The solid methane moderator was operated in phase II temperature and the energy spectra were measured at the temperatures of  20K and 4K.  We have also tested our newly-developed scattering kernels for phase II solid methane by calculating the neutron spectral intensity expected from the methane moderator at the LENS neutron source using MCNP (Monte Carlo N-particle Transport Code). Within the expected accuracy of our approximate approach, our model predicts both the neutron spectral intensity and the optimal thickness of the moderator at both temperatures.
The predictions are compared to the measured energy spectra. The simulations agree with the measurement data at both temperatures.
\end{abstract}

\maketitle

\section{\label{sec:level1}Introduction}
The Low Energy Neutron Source (LENS) at Indiana University Cyclotron Facility (IUCF) is a university based pulsed cold neutron source. It has been designed for education, research and neutron instrument development purposes. LENS produces neutrons from low energy ($p,~xn$) reactions in \textit{Be} target\cite{mark}.

LENS possesses a cold neutron moderator to convert the high energy neutrons from the \textit{Be} target to slow neutrons with energy spectrum and pulse characteristics suitable for  neutron scattering experiments in condensed matter\cite{watanabe}. The neutron energy is significantly reduced from $\sim eV$ to 10$meV\sim$ 0.1$meV$. The neutron slowing-down process is attained  by  scattering of epithermal neutrons on hydrogen  in the moderator\cite{yinwen}.
Solid methane is the brightest known moderating medium for pulsed cold neutron sources. This is because of its high hydrogen density and the rather unique presence of free rotor modes in the solid at low temperatures, which offers a mode of energy loss to the neutrons which possesses a relatively high neutron cross section and is not present in most other cold hydrogenous materials. Its use at high power spallation neutron sources is constrained by engineering difficulties associated with the radiation damage and hydrogen production in solid $CH_4$ in the intense radiation field near the spallation target \cite{carpenter1999}\cite{2002shabalin}. The radiation damage at lower power pulsed cold neutron sources can be low enough that one can operate the moderator at low temperatures for extended periods of time and at temperatures that are lower than practical at higher-power sources.
Normally, it is necessary at some point to warm and refreeze the moderator to release radiation damage energy stored in the lattice in a controlled way before a spontaneous recombination occurs which can produce high transient pressures and burst the moderator vessel. Nevertheless, most neutron source facilities which use solid methane as a moderator choose not to operate the moderator below 20K.  For temperatures below 20K, solid methane enters a phase (called phase II) in which only 1/4 of the rotational modes remain free while the remaining 3/4 of the modes undergo librations and tunneling motions. Since phase I possesses free rotor modes for all sites, it is not obvious that the colder spectrum that one might be able to achieve by operation of the moderator at lower temperatures can be realized in practice due to the loss in free rotor modes. It was observed that the effective temperature of the neutron spectrum, which is typically about a factor of two larger than the physical moderator temperature for most cold neutron moderators, showed stronger and stronger deviations from the physical temperature of the methane in phase II. This observation, coupled with the added inconvenience of repeatedly cycling the moderator through the solid-solid phase transition at 20K between phases I and II in the course of releasing the stored energy from radiation damage, accounts for the 20K operating temperature choice of most sources. Partly for this reason, existing neutron scattering kernels do not consider the modification of the neutron scattering dynamics which should occur in solid methane as a result of this phase transition since few moderators operate in this regime.

However, the LENS neutron source operates in a qualitatively different regime. Since it is based not on spallation nor on fission but rather uses $(p,xn)$ reactions in a \textit{Be} target and because the time-averaged neutron production is somewhat lower than typical reactor and spallation sources, it is practical to operate the moderator at much lower temperatures and with greatly reduced radiation damage effects compared to these sources. This possibility provides a strong motivation for operating a methane moderator at low temperatures well into the phase II regime. 

Parallel to the operation, important theoretical work had to be carried out both on the general formalism and on the question of finding a model which should describes approximately the physics of all the scattering processes in phase II and can be used for analyzing the experimental data. We have, therefore, developed an approximate model of $S(Q,\omega)$ for solid methane in phase II which is consistent with the known spectroscopy of the low energy modes and the measured neutron total cross section this neutron scattering model and relevant results have been reported\cite{yun2007-1}.

To see if this new scattering kernel can describe the neutron spectra from a realistic neutron source, the neutron spectra were not only obtained with the simulation at two temperatures but also measured from our LENS solid methane moderator using  neutron time-of-flight methods. In this paper, we describe our measurement method, present the experimental results for the  neutron spectrum measurements at temperatures of 4K and 20K, and compare the results with those obtained from the simulation in order to confirm the validity of our scattering kernels.

\section{\label{sec:level1} Solid Methane in phase II}

The ${CH}_{4}$ molecules in solid methane may be viewed as quantum mechanical spherical rotors at temperatures below $\sim20\textrm{K}$ (phase II). The methane in this phase consists of two sublattices. 1/4 of the molecules rotate almost freely. The lowest energy levels are at 0, 1.09 and 2.56 ${meV}$. The remaining 3/4 of the molecules librate in the deep minima of a strong orientational potential and form a threefold multiplet with levels of 0, 0.16 and 0.24${meV}$. The librational state energies are present at  6.5${meV}$ and above. As a consequence of the Pauli principle as applied to the 4 identical protons in the ${CH}_{4}$ molecule, each rotational state of symmetry \textit{A},\textit{T} and \textit{E} are associated with a definite total spin of the four protons, $I=2,1,0 $ respectively. Transitions between levels of different symmetry, which require a change in the spin state of the protons,  cannot be induced by phonon interactions alone but must be  mediated by spin-dependent interactions such as the weak dipole-dipole interaction among protons. As a consequence, the spin system shows very slow relaxation to thermal equilibrium  after a sudden change of the lattice temperature\cite{Friedrich:1996lr}.

\begin{figure}[htbp]
\begin{center}
\includegraphics[width=10cm]{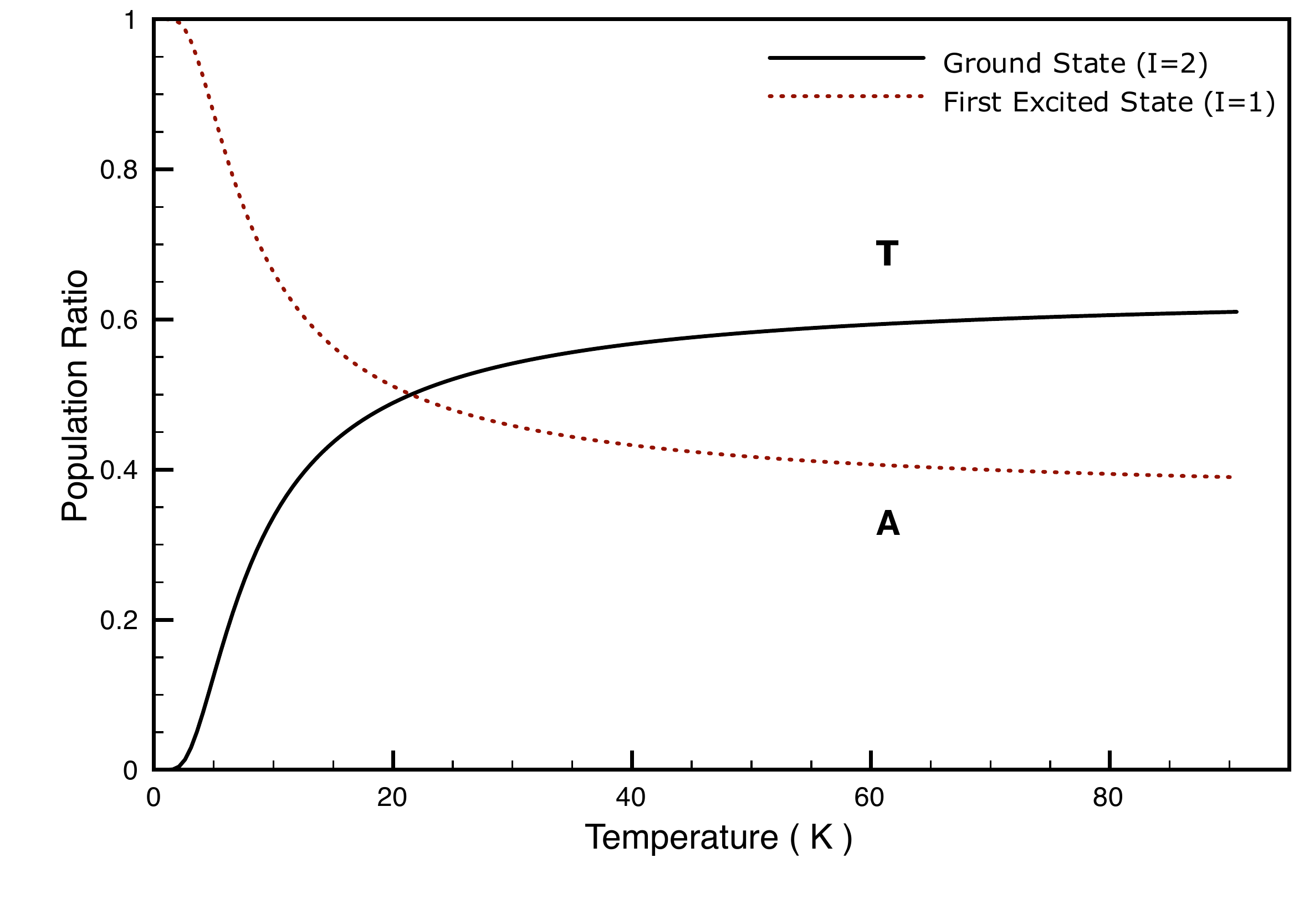}
\caption{The spin population in free rotation mode}
\label{fig:a}
\end{center}
\end{figure}

\begin{figure}[htbp]
\begin{center}
\includegraphics[width=10cm]{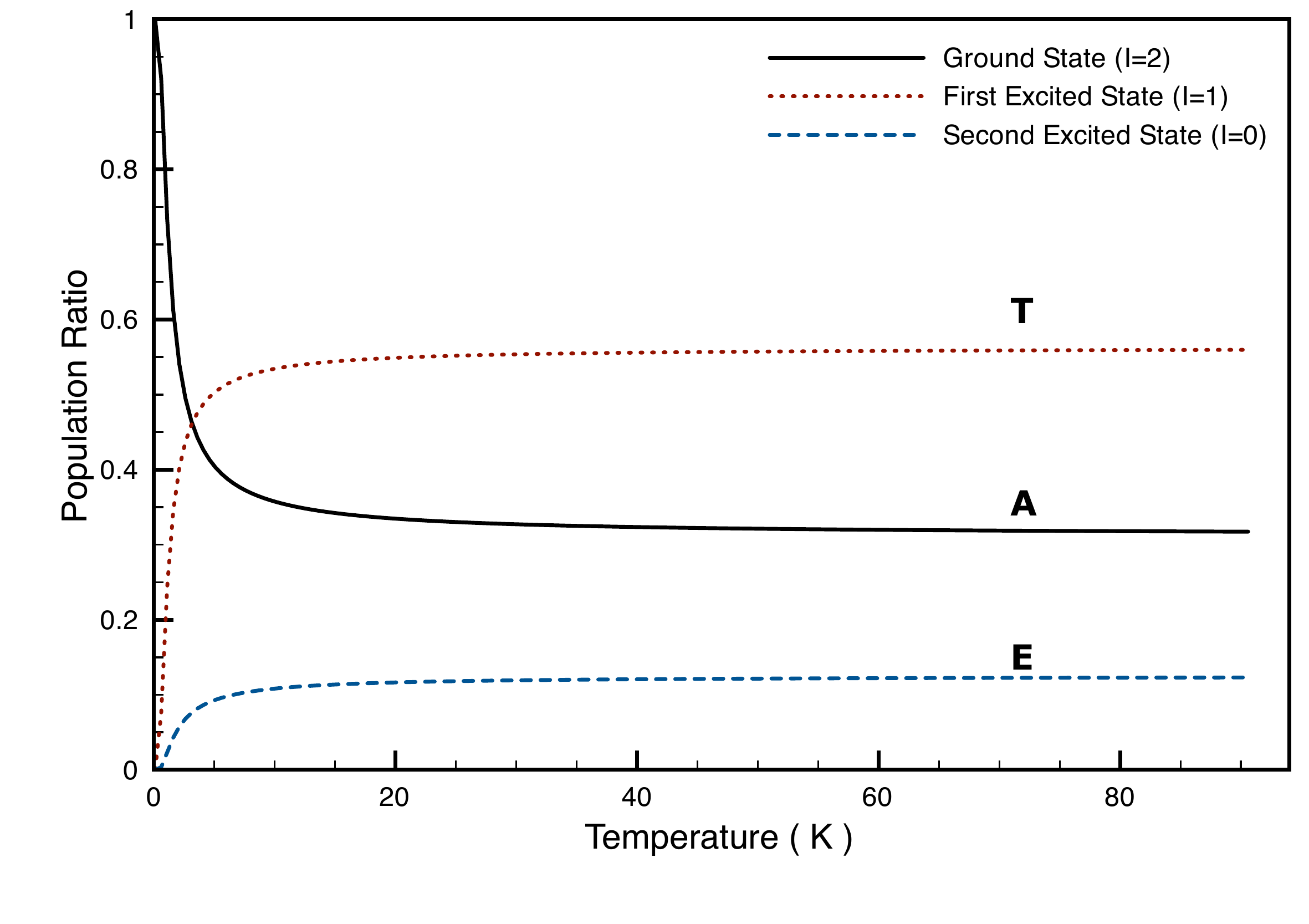}
\caption{The spin population of hindered rotation}
\label{fig:b}
\end{center}
\end{figure}

Fig.~\ref{fig:a} and \ref{fig:b} show the relative populations of the ${CH}_{4}$ molecules in the free rotation and hindered rotation groups. These spin populations are calculated assuming a Boltzmann distribution. Although the high temperature approximation is appropriate for methane at room temperature, the ratio doesn't change even after the melting point of solid methane ($\sim$ 90.6K). The  restriction to a two energy states system in free rotation and a three energy states in hindered rotation is justified at low temperature due to the weak population of higher excited states\cite{1994hepp}. The spin distribution function  is
\begin{equation}\label{eq:1.1}
P_{i}=g_{i}\frac{\mathrm{exp}(-E_{i}/k_{B}T)}{\displaystyle\sum_{i} \mathrm{exp}(-E_{i}/k_{B}T)}\end{equation}
where $g_{i}$ is occupation number of each state.

It has been noticed in several experiments that methane shows only slow relaxation of the spin system even when changing the temperature of the lattice. The theory of nuclear spin conversion was first developed for the ortho-para transition of hydrogen and later was extended to solid methane by Nijman $\&$ Berlinsky\cite{NIJMAN}. The importance of the spin-spin interaction has already been shown by Curl \textit{et~at.}\cite{CURL}. The effect of paramagnetic ${O}_{2}$ impurities has been discussed by Kim \textit{et~at.}\cite{KIM}.

N$\&$B have translated the conversion process into a microscopic view. Rates of the conversion from $T\rightarrow A$ symmetric states are evaluated using the Fermi golden rule,
\begin{equation}\
\alpha=\frac{1}{\tau}=\frac{2\pi}{\hbar}\sum_{i,f}P_{i}\mid\langle\Psi_{f}\mid H^{tr}\mid \Psi_{i}\rangle\mid^{2}\delta(E_{i}-E_{f}).
\end{equation}
The perturbation operation $H^{tr}$ couples different initial and final states $\mid\Psi_{i}\rangle$ and $\mid\Psi_{f}\rangle$. $E$ denotes the energy of the states and $P_{i}$ the relative occupation of the initial state. There have to be at least two different kinds of interactions to induce a conversion process from one spin species to another. For energy conservation, the coupling to phonons has to be considered. Additionally spin states have to be changed. This requires the presence of magnetic field gradients, i.e. unpaired electrons (e.g. ${O}_{2}$  impurities) or dipole-dipole interactions of protons\cite{Friedrich:1996lr}.

The calculation of the total neutron scattering cross section of rotating molecules have shown that
\begin{equation}
\sigma_{tot}\propto \langle I(I+1) \rangle~\textrm{with}\langle..\rangle :\textrm{thermal}~ \textrm{average}
\end{equation}
is a good approximation for long wavelength neutrons\cite{Friedrich:1996lr}.
Since each rotational state is related to a distinct total nuclear spin $I$, $\sigma_{tot}$ is proportional to their occupation number $g_{i}$ in Eq. \ref{eq:1.1}. $\sigma_{tot}$ increases with decreasing temperature because
$I$ is higher for the lowest spin levels. The higher temperature limit is determined purely by the number of protons. In the case of methane, it may include the scattering from free and ordered molecules,
\begin{equation}
\sigma_{tot}=\frac{1}{4}\sigma_{free}+\frac{3}{4}\sigma_{hindered}.
\end{equation}
Ozaki \textit{et~at.} have calculated the absolute value of the total neutron scattering cross section for methane in this phase\cite{ozaki:3442}.

Both the liquid-solid and solid-solid phase transition, which can lead to nonuniformities in the solid methane density,  and the sensitivity of the neutron scattering in methane to the spin state of the protons can cause difficulties in accurately predicting the neutron energy spectrum from a practical moderator. To mitigate these possible problems, we added a small concentration of $O_{2}$ paramagnetic impurity to boost the nuclear spin conversion rate and we cooled the moderator slowly in an attempt to avoid the cracks and holes in the moderator.  Cracks and holes in the moderator may appear during the thermal contraction of the moderator material which will change the macroscopic neutron cross section. The conversion process of spin species of both rotational system, free rotor and hindered rotational system, contribute to the total scattering cross section $\sigma_{tot}$.  In all cases, the conversion rate of two rotational systems are different. For temperature above $\sim$4K, the occupation number of the hindered molecules are always near their high temperature limit and consequently the conversion behavior is dominated by the free rotation mode\cite{Friedrich:1996lr}.

\section{\label{sec:measurement} The Neutron Spectrum Measurement}

\subsection{\label{sec:configuration} The Configuration for the Measurement}
\begin{figure}[htbp]
\begin{center}
\includegraphics[width=10cm]{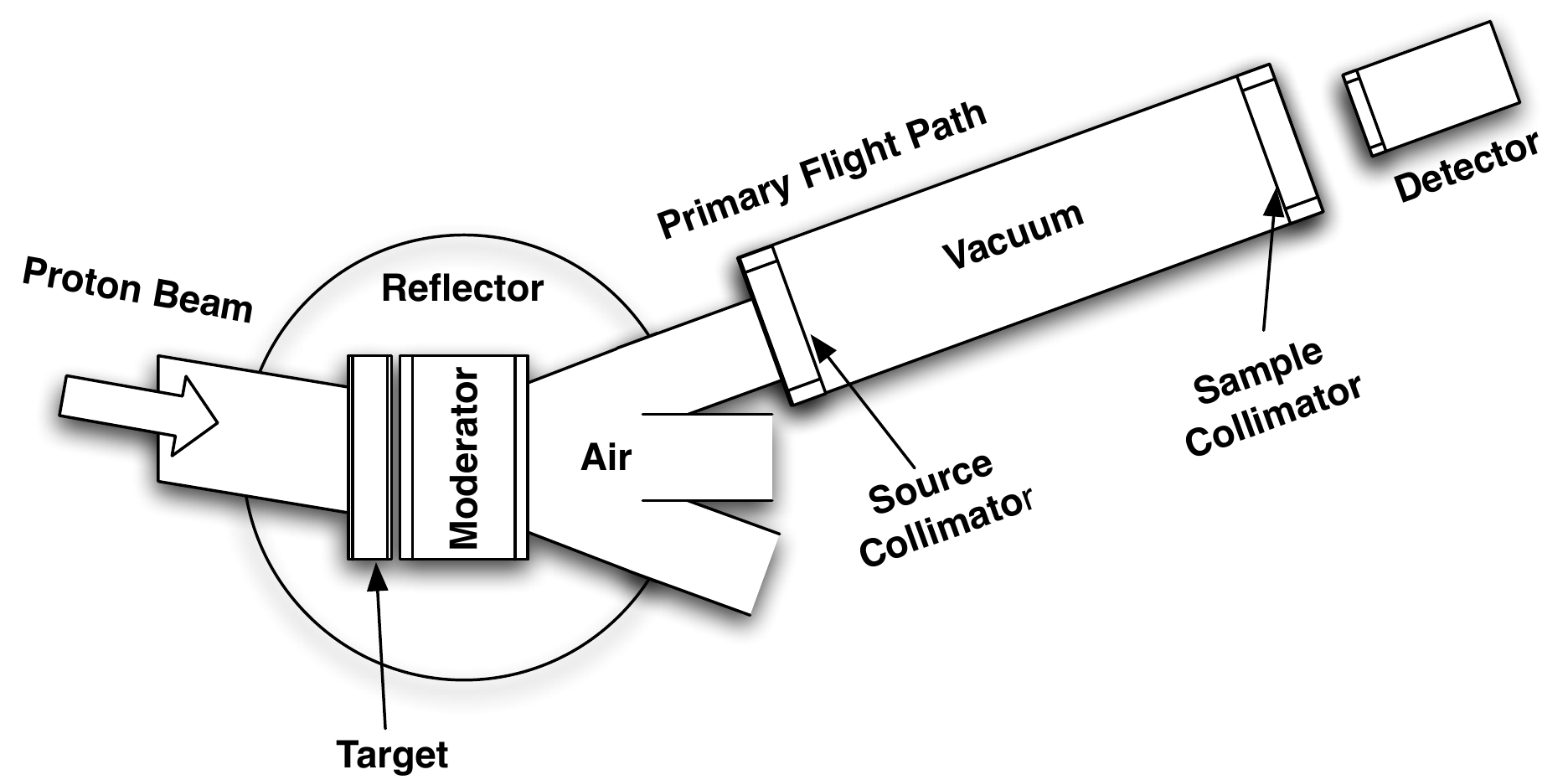}
\caption{The Systematic View of LENS beamline}
\label{fig: tmr}
\end{center}
\end{figure}

The LENS neutron source presently has three neutron beam lines for SANS, Neutron Radiography and instrument development. These beam lines are oriented at $-20^\circ$, $0^\circ$ and $20^\circ$ relative to the normal surface of the moderator. The neutron spectrum measurements were performed in the SANS beam line. Fig.~\ref{fig: tmr} shows the layout for the neutron spectrum measurement.

Neutrons were generated in the $Be$ target from $^{9}Be~(p, xn)$ nuclear reaction with protons.
The incident protons are introduced from a commercial liner accelerator with the energy of 7 \textit{MeV}. In the linac, protons have square-shaped pulse widths of 170$\mu s$. The peak current of the linac was 8.5 \textit{mA} and the frequency of the accelerator was chosen to be   $15Hz$ for these measurements. The \textit{Be}  target thickness was  chosen for its neutron yield and mechanical strength under steady-state and transient thermal stresses induced by the proton beam \cite{2007chris}. The proton beam of area $4\times4cm^{2}$  is incident on the target surface with an $45^\circ$ to reduce the power density on  the \textit{Be} target.
The moderator is placed right next to the target with a minimum vacuum space of $5.5cm$. The current design of TMR (Target, Moderator and Reflector) system has a room temperature light water reflector surrounding \textit{Be} target and the moderator. The area of the moderator is $12\times12cm^{2}$ and the thickness of $1cm$  was optimized using  MCNP model calculation based on the 22K solid methane scattering kernel ``\textit{smeth22K}". Since this optimization exhibited a broad maximum from $1cm$ to $3cm$, we decided to choose a moderator thickness on the low end of the distribution to minimize the total amount of matter and therefore the neutron and gamma heat load on the moderator.

We used a  5.6$m$ primary flight path for the energy spectrum measurements. The entrance beam window of the tube is located at 1.4$m$ from the moderator surface. The tube was evacuated to a pressure of 0.1 $mbar$. Two pinhole collimators were mounted at the ends of the tube to define the solid angle of the viewed moderator. A beam collimator of 3" diameter was mounted on the moderator side and the sample collimator with 1" diameter was on the detector side \cite{2004NIMPA.517..285S}.

The neutrons emitted from the moderator surface first pass through 1.4$m$ of air  and enter the vacuum tube through a single crystal $Si$ window of  5.82$mm$ thickness. The neutrons leave the vacuum tube through a second 5.82$mm$ thick $Si$ window  and were collimated by a pinhole with 1$cm$ diameter. A high efficiency $^{3}He$ detector pressurized to 10 $atm$ was placed 0.05$m$ next to the primary flight path to scan neutron counts. The total neutron flight length was 5.65$m$. The 7$MeV$ proton linac trigger signal was used to define the zero time for the time-of-flight measurement of the neutrons.

\subsection{\label{sec:cooling} The Moderator Cooling}
We measured  the neutron energy spectrum  for the solid methane moderator held at temperatures of 20K and 4K. The moderator vessel is made with high purity aluminum and is  anchored to a high-purity aluminum rod for cooling with  a mechanical refrigerator. The moderator vessel was connected to the gas handling system through a gas feed line  wrapped with a 50$W$wire heater to prevent freezing of the methane.  Methane of $99.99\%$ purity was condensed from  room temperature  with an addition of $1\%$ of $O_{2}$ gas to ensure the neutron spin relaxation in the solid phase. The temperature of the moderator was controlled with a wire heater and an additional 50$W$ heater on the rod and monitored with two calibrated $Cernox$ temperature sensors which are directly attached to the top and bottom of the cell. During the measurement, the temperature gradient between the two sensors was less than 0.1K.

To be successful in  minimizing possible cracks and holes in cooling and solidifying the solid methane in the moderator vessel, we followed the general principles of the single crystal growth. The moderator vessel was first evacuated to $10^{-6}mbar$ at room temperature and cooled to 93K, which is just above  the crystallization temperature of solid methane.  The methane gas was liquified at this temperature with an  0.5 $mbar$ over pressure. After condensation, the liquid methane was cooled  toward the solidification point at a rate  of 1K/15min. At the crystallization temperature, the methane was annealed for 4 hours. The annealed methane was then cooled  to 54.1K. The average cooling rate in this step was about 0.5K/min. Then the methane was cooled again toward the solid-solid phase transition temperature in methane with same cooling rate and held at  20.4K for 3 hours to complete the phase transformation.  The moderator was then cooled to 20K to measure the neutron spectrum. The neutron spectrum measurement at 4K was performed  after we cooled the moderator at a  0.1K/min rate and held the moderator at 4K for 3 hours.

\subsection{\label{sec:analysis} Data Analysis}

\subsubsection{\label{sec:mean}Mean Emission Time Correction}
 We want to extract the neutron energy spectrum from the observed time-of-flight measurement. Due to effects such as the finite geometry of the moderator and statistical fluctuations in the moderation process, it is possible for a neutron emitted from the moderator with one  energy to overtake a slightly slower neutron that was emitted earlier. Therefore neutrons with different energies can arrive at the detector at the same time, which complicates the conversion of time-of-flight information to neutron energy  \cite{1963koppel}\cite{1970parks}.

The neutron energy spectra were calculated from the following formula,
\begin{equation}\label{eq:4}
C(t)=A \int_{0}^{t} dt\int_{0}^{\infty} dE\phi(E,t)\varepsilon(E),
\end{equation}
where $\phi(E,t)$ is the energy spectrum at time $t$ and $\varepsilon(E)$ is the efficiency of the detector.The number of counts accumulated in each time channel $t_{0}$ is
\begin{equation}\label{eq:10}
\frac{dC(t_{0})}{dt_{0}}=A\int_{0}^{\infty}\phi(E,t)\varepsilon(E) dE,
\end{equation}
where $C(t_{0})$ is the total number of count in the time interval $0<t< t_{0}$ with $t=t_{0}-\frac{L}{v}$. In the ideal case in which the flight path is very long and the emission time distribution from the moderator is narrow one might write
\begin{equation}\label{eq:6}
\phi(E,t)=\phi(E)\delta(t_{0}-\frac{L}{v}).
\end{equation}
If we assume that all neutrons of a given energy are emitted from the assembly at the same instant, this equation can still hold with,
\begin{equation}\label{eq:7}
\phi(E,t)=\phi(E)\delta(t_{0}-\tau_{a}(E)),
\end{equation}
where $\tau_{a}(E)$ is the emission time. The emission time can be defined as
\begin{equation}\label{eq:8}
\tau_{a}(E)=\frac{\int\phi(E,t) t dt}{\int\phi(E,t) dt}.
\end{equation}
We relied on an MCNP simulation to determine  the mean emission time distribution from our methane moderator since the neutron intensity at  LENS during this measurement was insufficient to determine it experimentally \cite{2007chris}.

When we substitute Eq.~\ref{eq:7} into Eq.~\ref{eq:10} and integrate, we find
\begin{equation}\label{eq:12}
\frac{dC(t_{0})}{dt_{0}}=A\phi(E)\varepsilon(E)\frac{2E}{\tilde{t}}\left|{1-\frac{d\tau_{a}(E)}{dt}}
\right|_{t=t_{0}},
\end{equation}
where $\tilde{t}=t_{0}-\tau_{a}(E)$. In the Eq.\ref{eq:12}, $dt_{0}$ is now the unit time $\Delta t_{0}$ and $dC(t_{0})$ is the counting rate $N(\Delta t_{0})$ in the unit time with a unit of $n/s$. $N(\Delta t_{0})$ should be background subtracted and normalized. This equation now includes the correction for the effect of finite emission time. Then, the neutron flux $\phi(E)$ is
\begin{equation}\label{eq:11}
\phi(E)=\frac{N}{\Delta t_{0}A\varepsilon(E) \frac{2E}{\tilde{t}}\left|{1-\frac{d\tau_{a}(E)}{dt_{0}}}
\right|}
\end{equation}
with the unit in $n/cm^{2}/s/meV$.
The quantity we used in the paper for the comparison is the neutron energy spectrum $I(E)$ from the following relationship,
\begin{equation}
E\times I(E)=\frac{L^{2}}{i_{p}}E\times\phi(E),
\end{equation}
where the $i_{p}$ is averaged the proton current during the time interval. The neutron energy spectrum $I(E)$ has unit of $n/meV/\mu C/sr$.

\subsubsection{\label{sec:correction} Data Corrections}
The neutron energy spectra were corrected for the effects of attenuation by the 1.4$m$ length of air gap, the two 5.82 $mm$  $Si$ windows, the 0.5 $mm$ of stainless steel cylindrical detector body and  for the efficiency of $^{3}He$ detector.

The first correction was the attenuation caused by the air gap. Between the moderator and the vacuum tube, there was 1.4$m$ length of flight pass filled with 1 $atm$ air which also attenuate the neutron intensity. We assumed that the air contains roughly 78.1$\%$ of Nitrogen, 20.96$\%$ of Oxygen and 0.94$\%$ of Argon in 1 $atm$. The attenuation due to the air varies from 11$\%$ to 3.8$\%$ for 0.1 $meV$ $\sim$ 1$eV$ respectively.

The 4.2$m$ tube was evacuated  with two mechanical pumps during the experiment. In addition, it had two single crystal $Si$ windows on both sides which had 5.82 $mm$ thickness. The correction for the $Si$ windows was obtained from an effective cross section approximation.

The use of large single crystals of various materials like $Si$  as a filter for thermal neutron beam has long been known\cite{egelstaff-1}.  At high neutron energies, greater than about 1$eV$, the total neutron cross-section $\sigma_{t}$ of each of the materials is in the range of a few barns. But, at lower thermal energies, less than 0.1$eV$ where much of the coherent ``\textit{Bragg}" scattering is disallowed. Then, the effective cross-section for the single crystal is much reduced to the several contributions: absorption $\sigma_{a}$, incoherent $\sigma_{inc}$, coherent inelastic $\sigma_{inel}$ and any residual ``\textit{Bragg}" scattering $\sigma_{el}$. The effective cross section is thus determined by \cite{1980RScI...51.1299N}
\begin{equation}\label{eq:5}
\sigma_{eff}=(\sigma_{a}+\sigma_{inc}+\sigma_{inel}+\sigma_{el}).
\end{equation}

In Table~\ref{table:1},  we list the relevant cross section of the material we used for the correction. The values of $\sigma_{a}$ refers to $1.8\AA$ (25$meV$) neutrons.
\begin{table}[h!b!p!]
\begin{tabular}{|c|c|c|c|c|}
\hline
Element & Absorption & Incoherent&Coherent&Total \\
\hline
Al &0.230 & 0.1 &1.469&1.43\\
Si & 0.16 &0.02& 2.163& 2.25\\
Fe & 2.56 &0.4&11.22& 11.62\\
\hline
\end{tabular}
\caption{Cross Sections of Elements ($b=10^{-28}m^{2}$)}
\label{table:1}
\end{table}

All these cross-sections are reasonably well known quantities and are independent of orientation, perfection and temperature of crystals.  But the best possible effective cross section is obtained when the contributions of $\sigma_{inel}$ and $\sigma_{el}$ are eliminated \cite{1980RScI...51.1299N},
\begin{equation}\label{eq:6}
\sigma_{eff}=\left(\sigma_{a}+\sigma_{inc}\right)
\end{equation}

The attenuation due to these windows ranges from 12 $\%$ to 6.8 $\%$ for  0.1 $meV$ $\sim$ 1$eV$ respectively.

The attenuation for the 0.5 $mm$ thickness of stainless steel detector body was about 10.5$\%$ to
2.5$\%$ for  0.1 $meV$ $\sim$ 1$eV$ respectively.

The efficiency of the neutron detector depends on $^{3}He$ absorption cross section through the reaction
\begin{equation}\label{eq:1}
^{3}_{2}He + ^{1}_{0}n \rightarrow ^{3}_{1}He + ^{1}_{1}p, ~Q=0.764 MeV.
\end{equation}
The thermal neutron cross section for this reaction is 5330 barns for a 25 \textit{meV} neutron. The detector efficiency for the neutron incident along the $^{3}He$ detector tube is approximately given by
\begin{equation}\label{eq:2}
\varepsilon(E)=1-\mathrm{exp}(-n\sigma_{a} l),
\end{equation}
where \textit{n} is the number density of the atom, $\sigma_{a}$ is absorption cross section of $^{3}He$ at energy $E$ and $l$ is active length of the detector tube. Using Eq.~\ref{eq:2} we find the calculated efficiency for  the 1 $cm$ diameter tube filled with 10 $atm$ of $^{3}He$ is 99.14$\%$ at thermal neutron energy (25meV) but drops to 3.9$\%$ at 1$eV$. Thus a $^{3}He$ tube exposed to neutrons with various energies responded principally to slow neutron component.
Eq.~\ref{eq:2} slightly over estimates the neutron counting efficiency because there usually are some regions near the end of the tube in which charge collection is inefficient. It results in reduced neutron response. This dead space was considered in the efficiency correction \cite{1980knoll}

\subsubsection{\label{sec:level2} Energy Resolution}

In time-of-flight measurements, the distortion of the spectrum is mainly caused by the imperfect energy resolution. This imperfect resolution comes  from the uncertainty of the flight path length ($\delta L$), the finite width of the time channel ($\delta t_{c}$) and from the imprecise determination of the zero time when neutrons start to run across the flight path ($\delta t_{f}$). By definition, the imperfect resolution is equivalent to an uncertainty of neutron energy\cite{1970parks}.

For a pulsed neutron source, the finite pulse width of the proton beam and the finite time that neutrons spend in the moderator introduce  uncertainty. The  finite time results in the uncertainty of the zero time. As a consequence, the experimental energy resolution in the pulsed source is primarily limited by the average emission time $\tau_{a}(E)$ of neutrons and the slowing down time which broaden out the neutron burst.

The energy resolution arising from the finite channel width, the uncertainties in the flight-path length and uncertainty in the flight time is \cite{1970parks}
\begin{equation}\label{eq:3}
\delta E\simeq \frac{1}{2}E\left(\left(\frac{2\delta t_{f}}{t_{f}}\right)^{2}+\left(\frac{2\delta t_{c}}{t_{f}}\right)^{2}+\left(\frac{2\delta L}{L}\right)^{2}\right)^{1/2},
\end{equation}
where $\delta t_{f}$ is the $\tau_{a}(E)$ and $t_{f}$ is flight time\cite{1970parks}. For the high efficiency detector,  $\delta L=\lambda (E)$, the energy-dependent absorption mean free path of neutrons in the detector. $\delta t_{c}$ is the time channel width $\Delta t_{0}$. The correction for the uncertainty of the energy resolution improved the neutrons spectrum especially with energy less than 2$meV$.

\section{\label{sec:4-5} Neutron Energy Spectrum Calculation}

For the neutron spectra calculation in solid methane, the MCNP neutron transport code was used. The MCNP Scattering kernels were produced using NJOY nuclear data processing system with the frequency spectrum for solid methane at 20K and 4K from our model\cite{yun2007-1}.
We named these kernels as ``\textit{y-smeth20K}'' and ``\textit{y-smeth4K}". The geometry for MCNP simulation was the same as the experimental one in the LENS (Low Energy Neutron Source) beam line. A systematic view of the TMR (Target-Moderator-Reflector) is shown in Fig. \ref{fig: tmr}.  The neutron energy spectra emitted from the moderator was evaluated by point tally.
\begin{figure}[htbp]
\centering
\subfigure[20K] 
{
    \label{fig:4:a}
    \includegraphics[width=10cm]{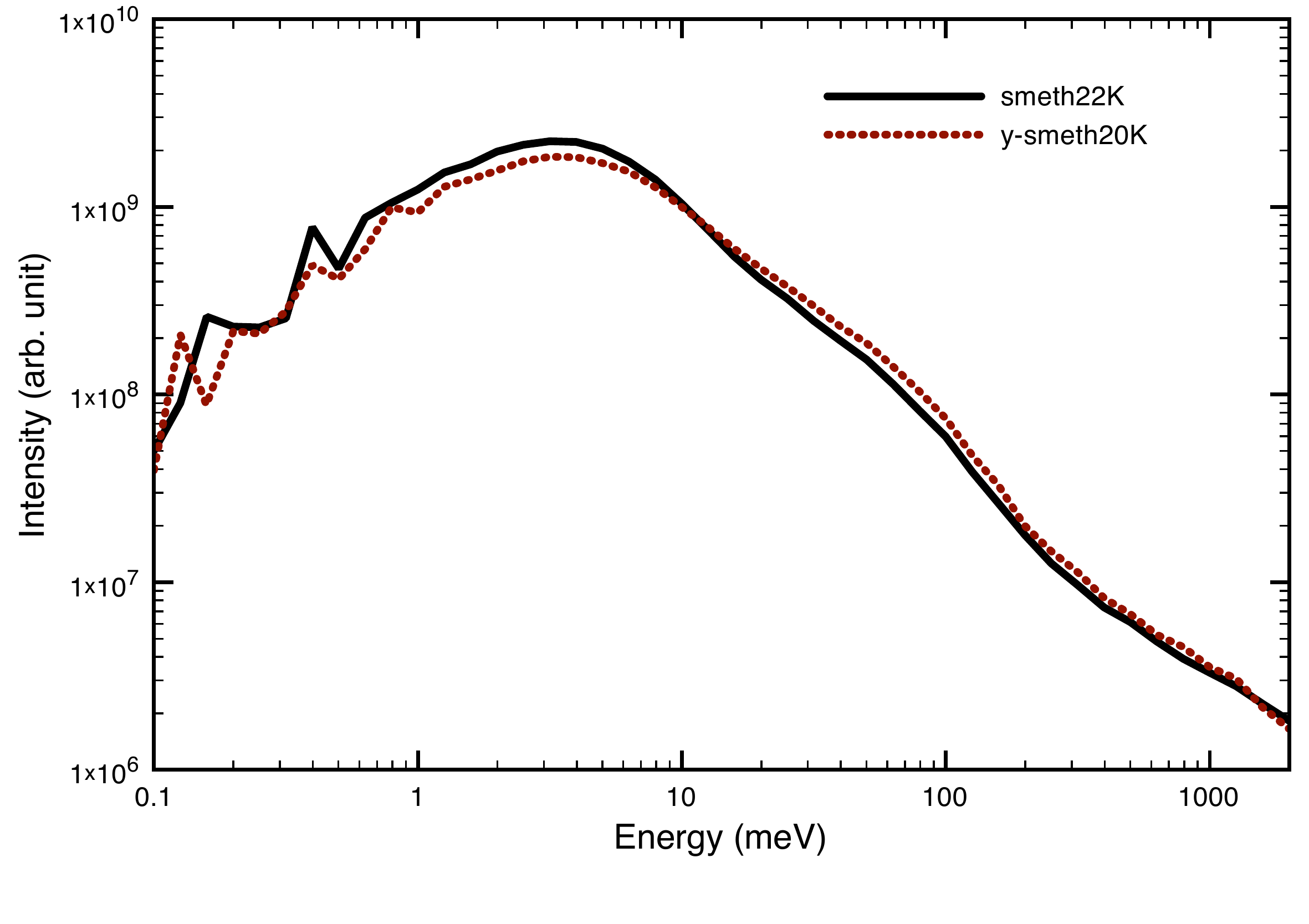}
}
\hspace{1cm}
\subfigure[4K] 
{
    \label{fig:4:b}
    \includegraphics[width=10cm]{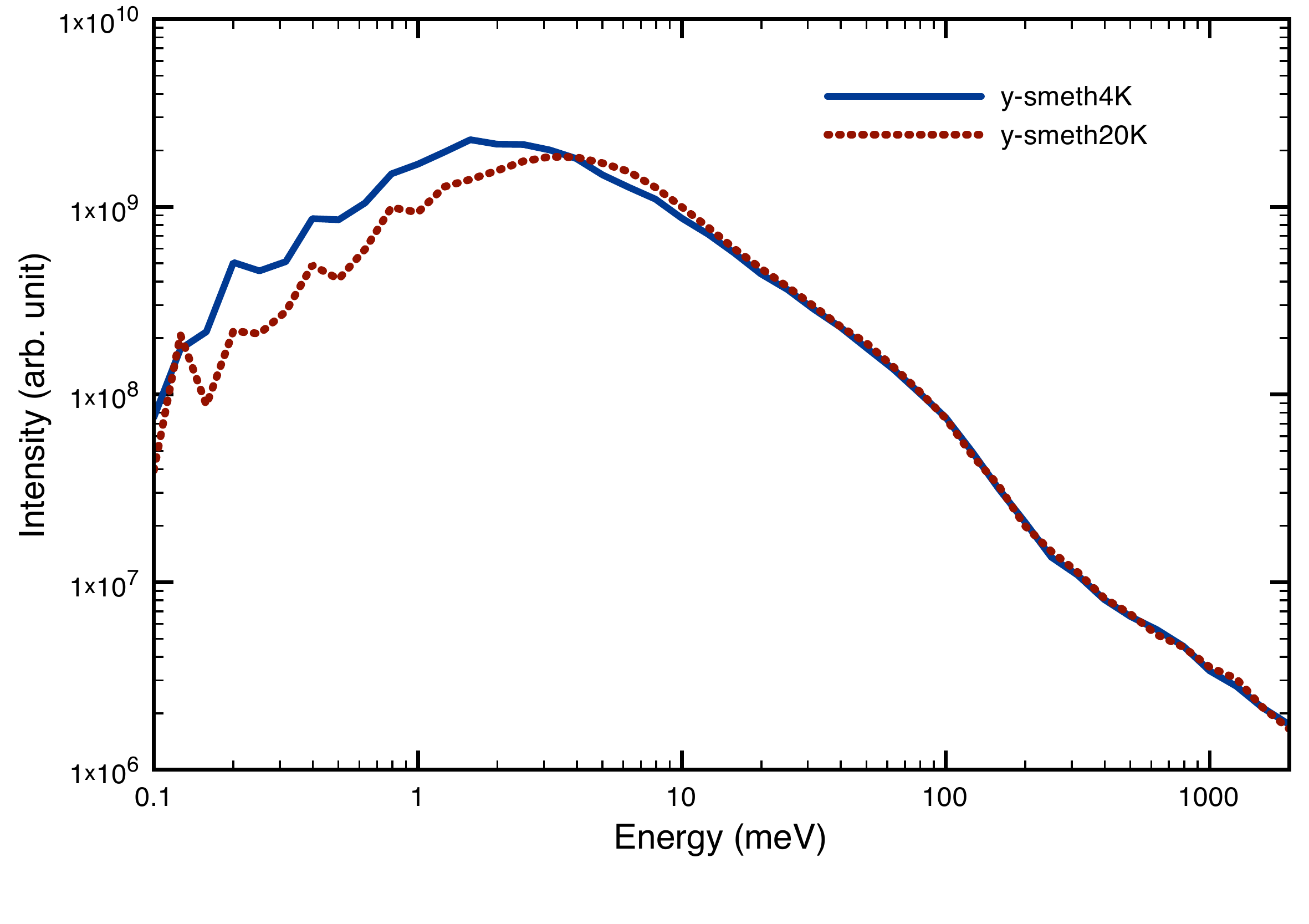}
}
\caption{The energy spectra of solid methane in 20K and 4K with 1cm moderator thickness.}
\label{fig:4} 
\end{figure}
Fiq. \ref{fig:4:a} and \ref{fig:4:b} show the calculated neutron energy intensity in 20K and 4K.  In the 20K plot, the neutron energy spectrum calculated with MCNP scattering kernel in 22K (\textit{smeth22k}) from Harker $\&$ Brugger frequency spectrum was also shown. The calculation with the 22K scattering kernel is close to our 20K kernel although there is a difference in neutron energy below 100 \textit{meV}.
\begin{figure}[htbp]
\centering
\subfigure[20K] 
{
    \label{fig:7:a}
    \includegraphics[width=10cm]{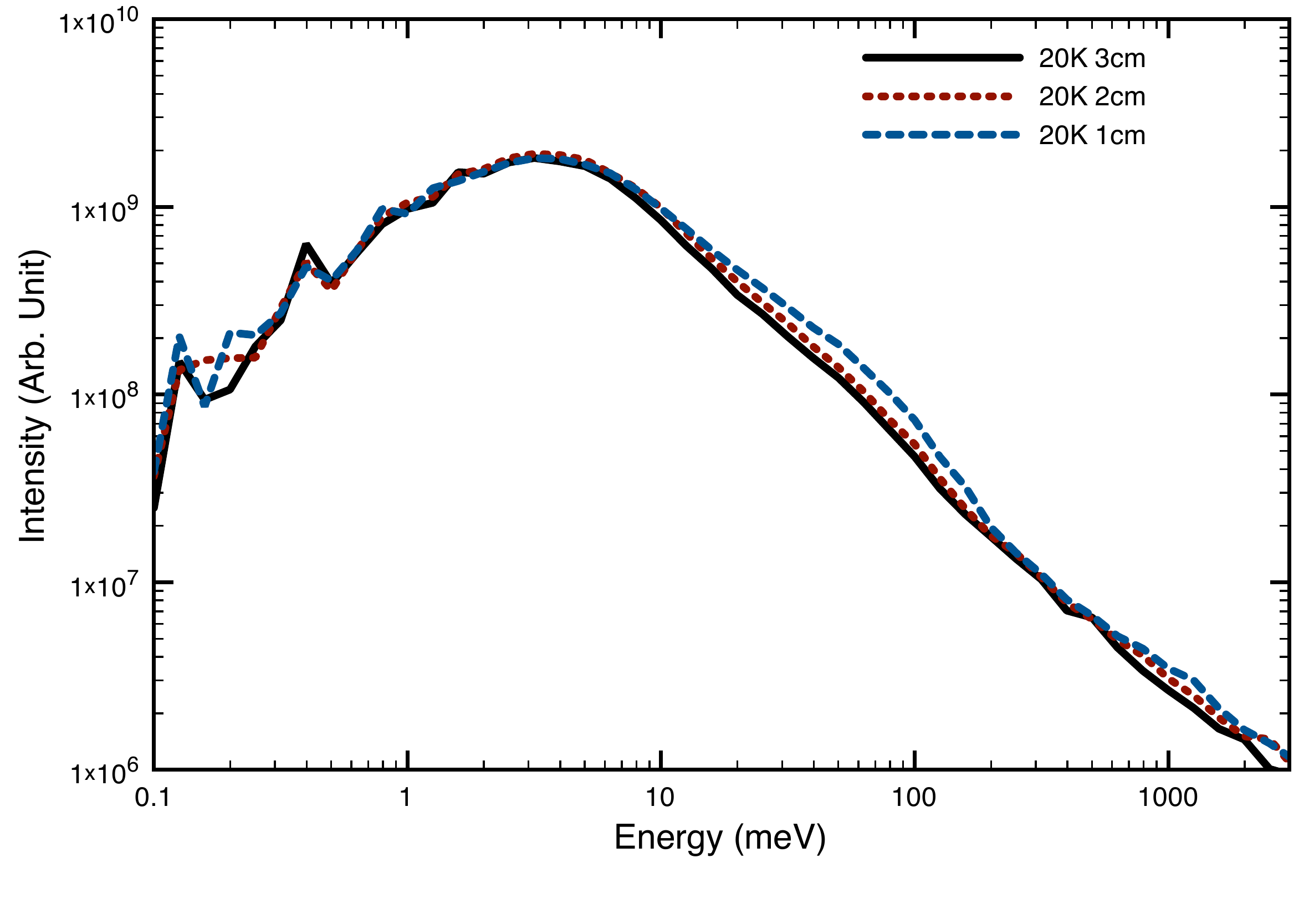}
}
\hspace{1cm}
\subfigure[4K] 
{
    \label{fig:7:b}
    \includegraphics[width=10cm]{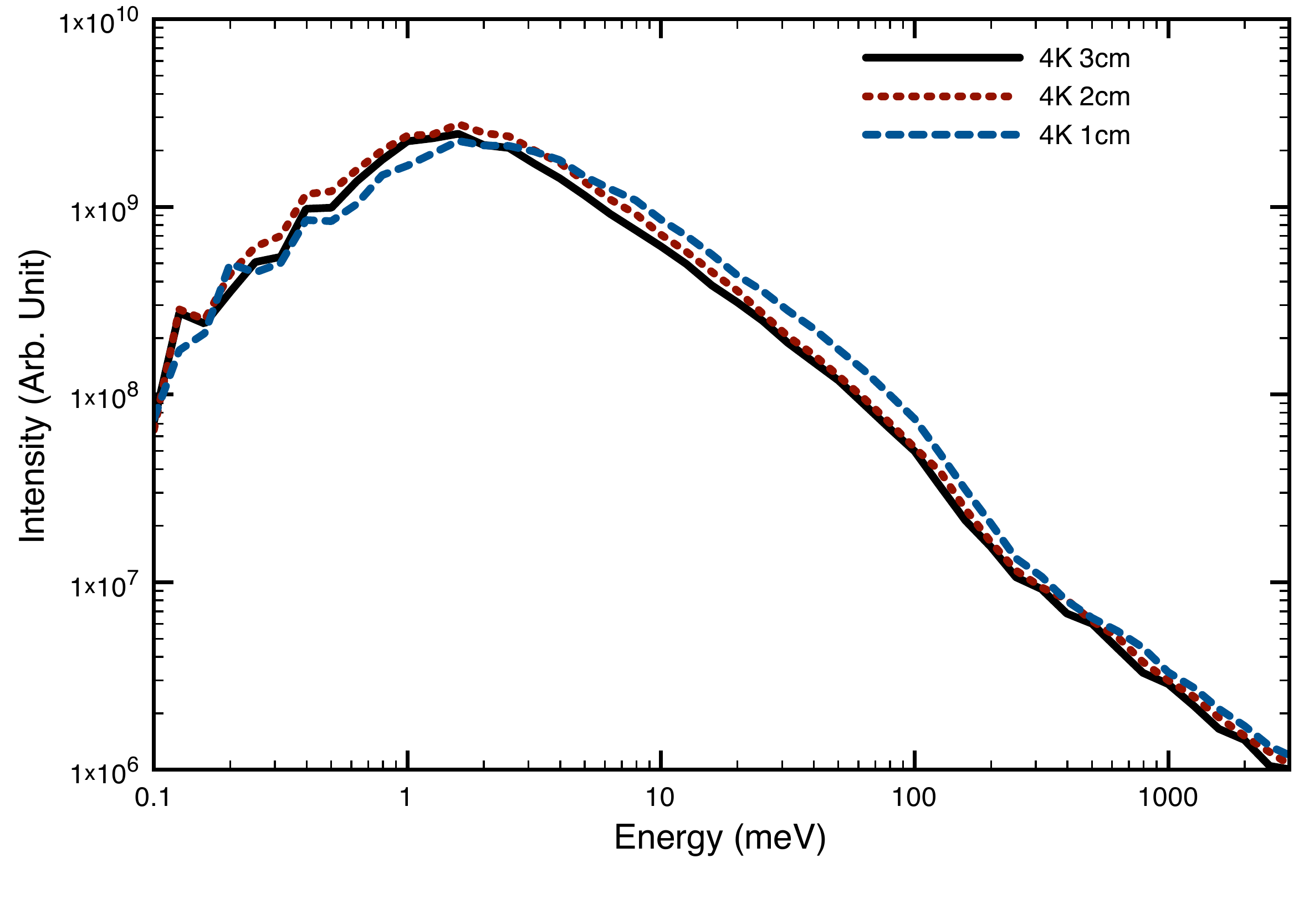}
}
\caption{The energy spectra of solid methane in 20K and 4K with various moderator thickness.}
\label{fig:7} 
\end{figure}

We also use these kernels to investigate the relationship of neutron spectrum to  the moderator thickness. We had increased the moderator thickness from 1\textit{cm} to 2\textit{cm} and then to 3\textit{cm} and performed MCNP simulation with our 20K and 4K kernels. In the 20K, the cold spectrum wasn't shifted to a lower energy region even though the thickness of the moderator had been increased up to 3\textit{cm}. At the same time, the intensity of thermal neutron had been reduced by about 35$\%$. This result implies  that   the moderator with 1\textit{cm}m thickness is the optimal thickness in the 20K moderator temperature with our LENS TMR configuration.

In the 4K shown in Fig. \ref{fig:7:b}, as the moderator thickness had been increased to 2\textit{cm}, not only is the intensity of thermal neutron reduced, but the cold neutron spectrum as well as it's peck point also is shifted to lower energy region. With the thickness of 3\textit{cm}, the overall intensity of the spectrum has been less than with 2\textit{cm} thickness. This results infer that about 20$\%$ of more intense cold neutron can be expected in the 4K moderator temperature with increasing moderator thickness to 2\textit{cm}, while the cold spectrum in the 20K is less changed.

\begin{figure}[htbp]
\begin{center}
\includegraphics[width=10cm]{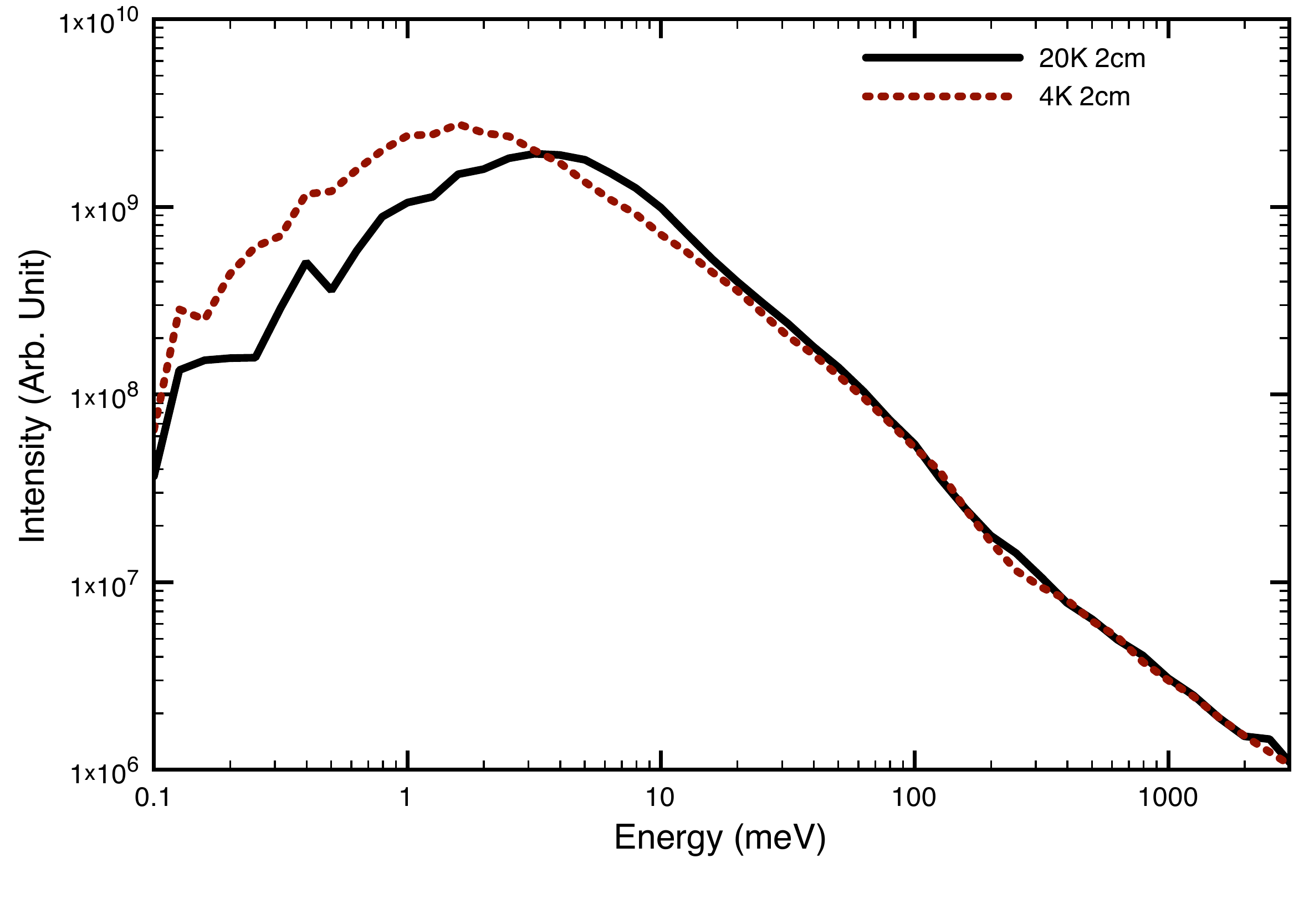}
\caption{The neutron energy spectra with 2cm moderator thickness}
\label{fig:9}
\end{center}
\end{figure}

\section{\label{sec:result} Results and Discussion}

\begin{figure}[htbp]
\centering
\subfigure[20K] 
{
    \label{fig:flux-a}
    \includegraphics[width=10cm]{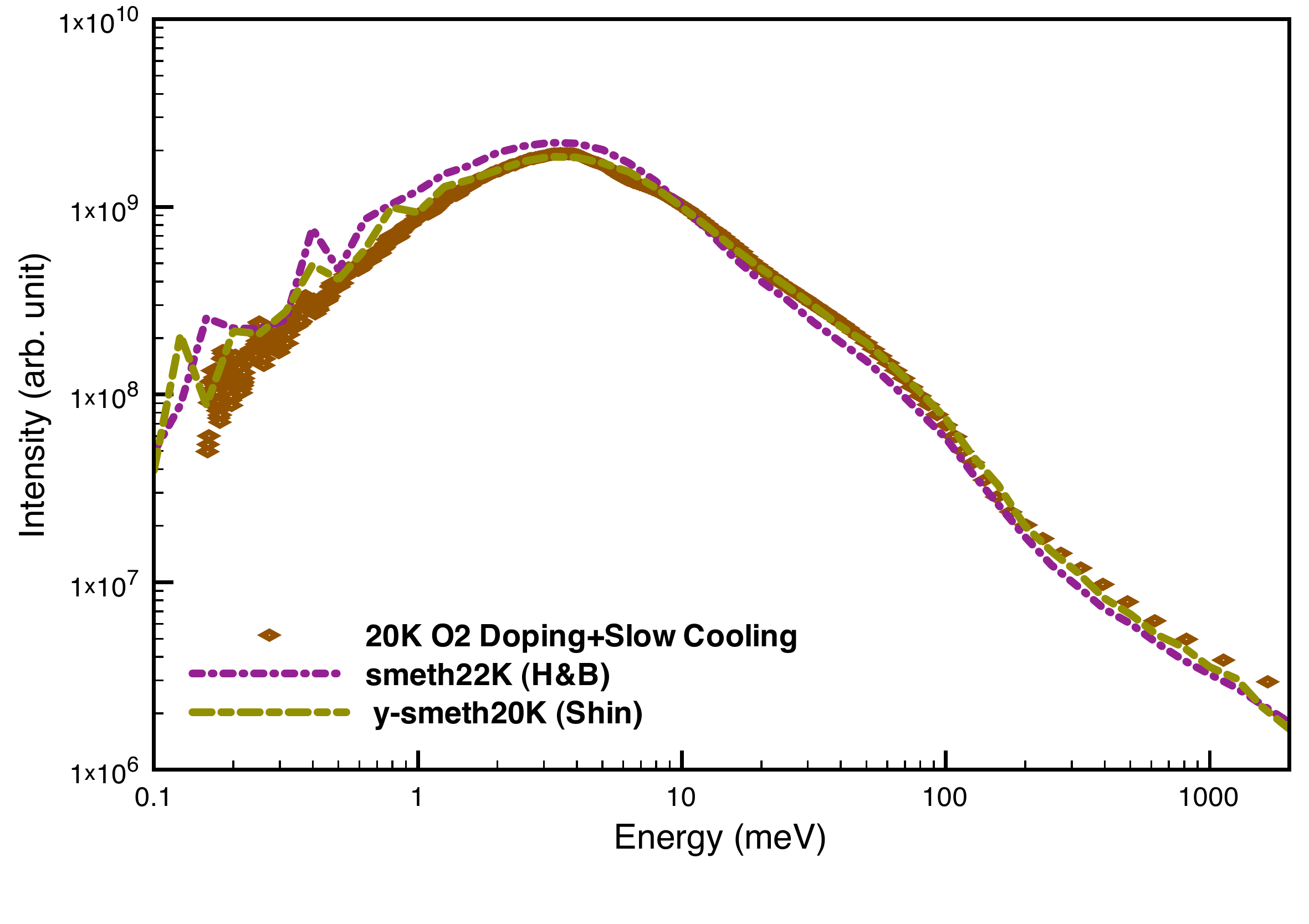}
}
\hspace{1cm}
\subfigure[4K] 
{
    \label{fig:flux-b}
    \includegraphics[width=10cm]{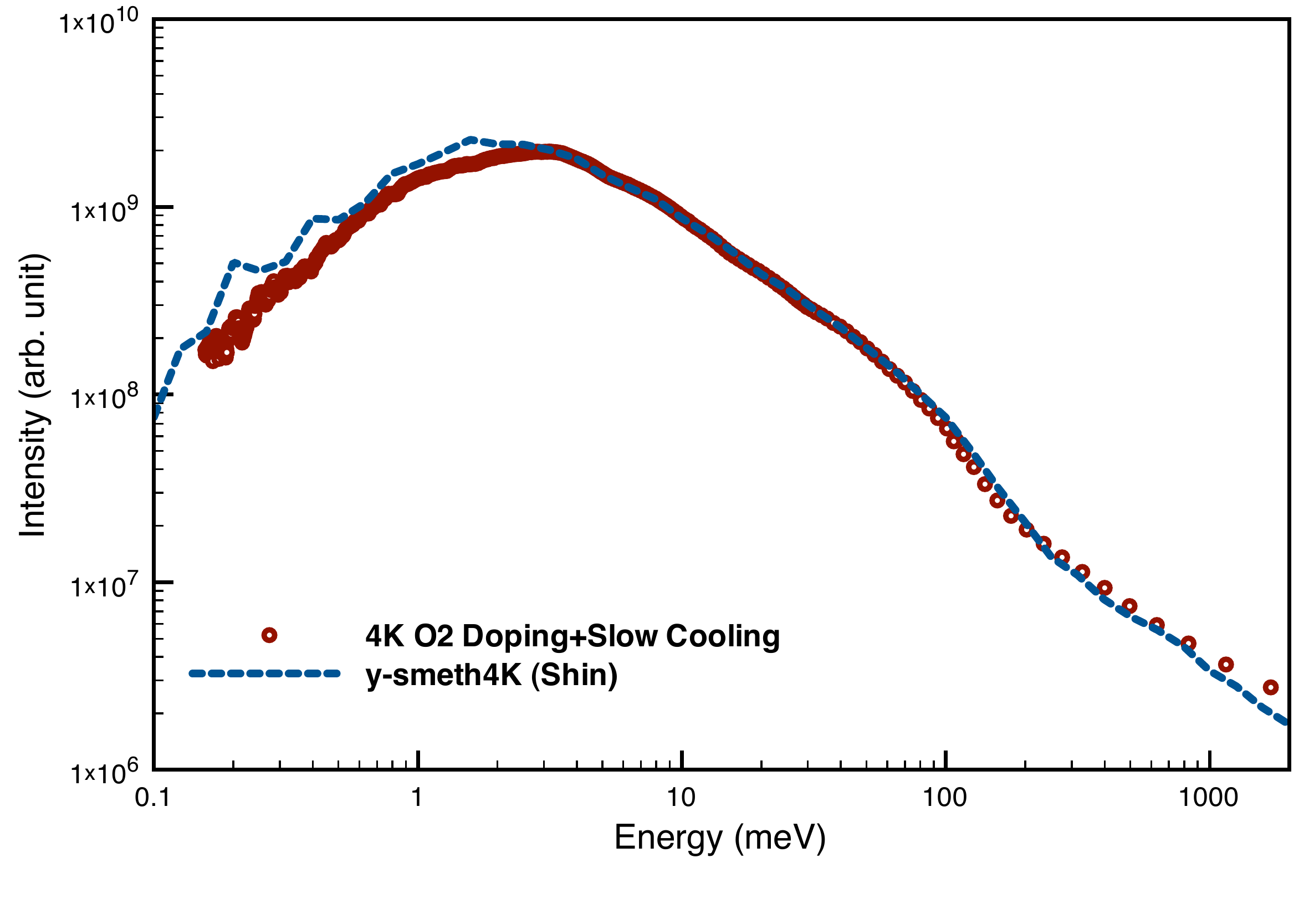}
}
\caption{The neutron spectrum in the 20K and 4K }
\label{fig:flux} 
\end{figure}

Fig. \ref{fig:flux} shows the neutron energy spectra from the solid methane moderator in the LENS system. The neutron energy range was measured in 0.1$meV\sim$ 1$eV$. The simulations were compared to the measurements at  20K and 4K. 

 The measured neutron energy spectrum at 20K  shows very good agreement with the MCNP simulation using our ``\textit{y-smeth20K}" kernel throughout the neutron energy range. The MCNP simulation with ``\textit{smeth22K}" kernel overestimates the measured intensity in the spectrum in the energy range below 10$meV$. 
 
The measured neutron energy spectrum at 4K agrees with the  MCNP simulation with our "\textit{y-smeth4K}" kernel in the energy range above 5$meV$.   However, the measured neutron spectrum is about 15$\%$ lower than the simulation  in the neutron energy across 2$meV$. 

\begin{figure}
\begin{center}
\includegraphics[width=10cm]{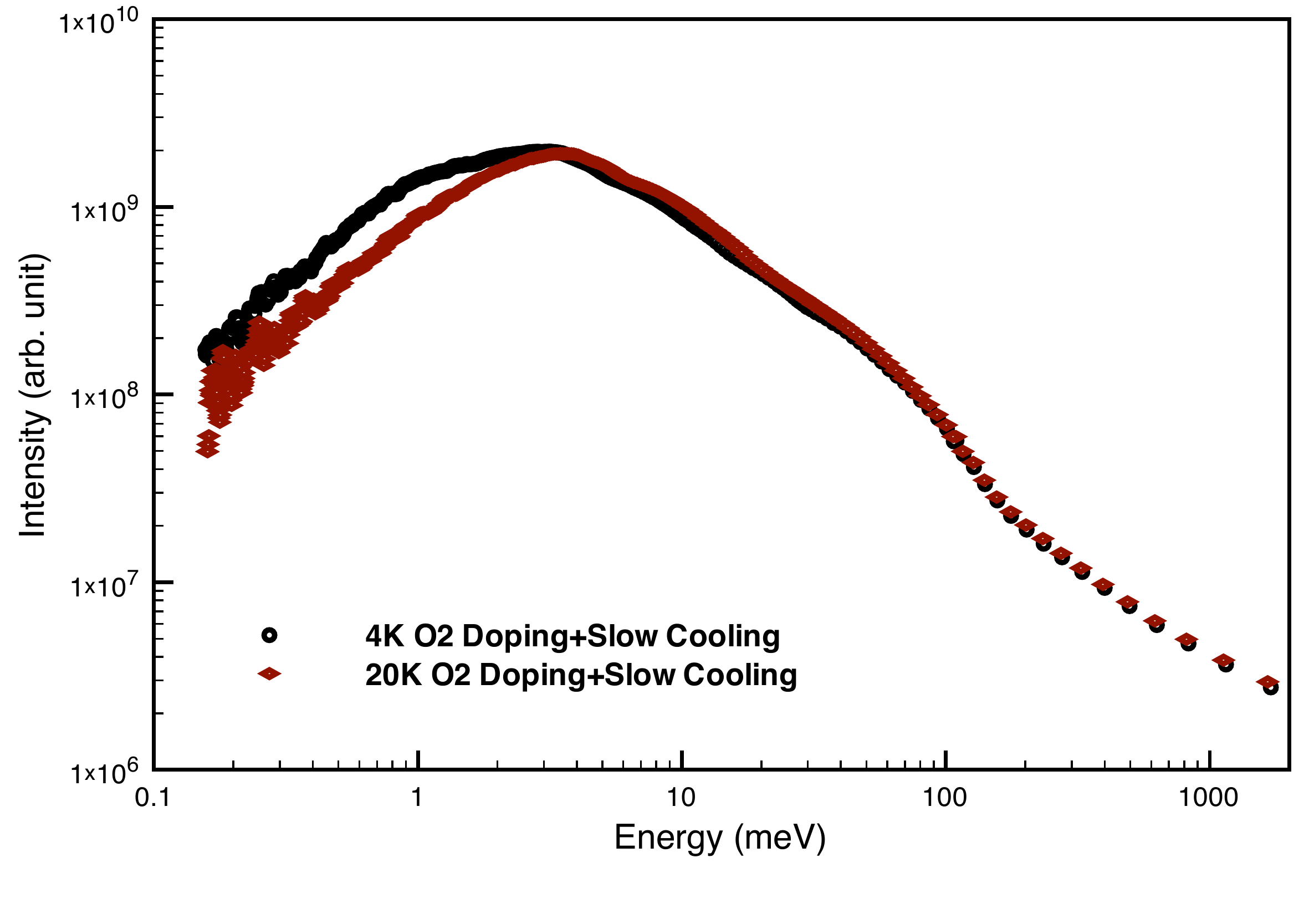}
\caption{ Neutron energy spectrum in the 20K and 4K}
\label{flux20K4K }
\end{center}
\end{figure}

Fig.~\ref{flux20K4K } is the comparison of the measured neutron spectrum in the 20K and 4K. It is clear that the 4K measured spectrum lies below the theoretical prediction  around 2$meV$ neutron energy.

\begin{figure}
\begin{center}
\includegraphics[width=10cm]{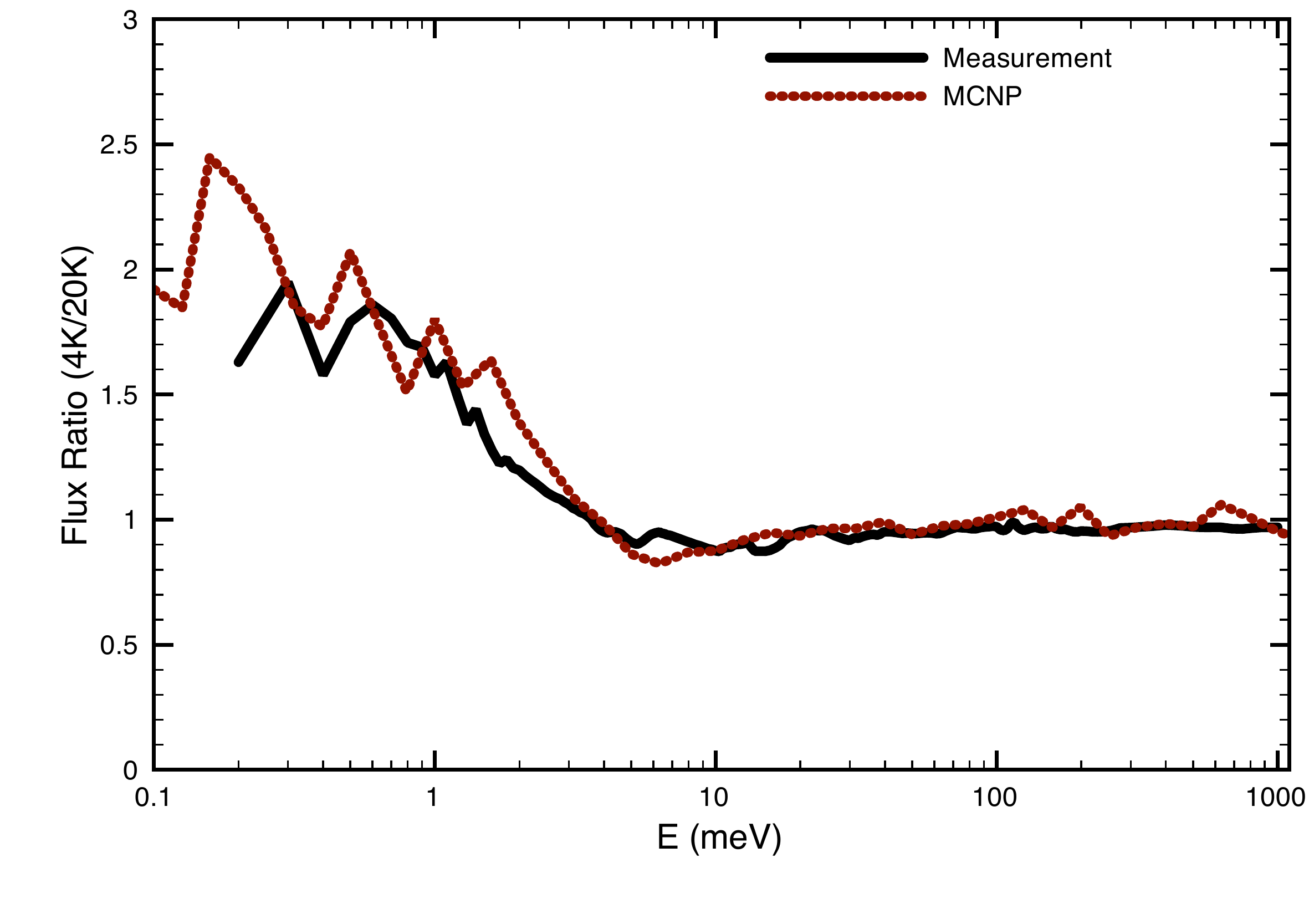}
\caption{ The ratio of 4K and 20K neutron energy spectrum }
\label{ratio}
\end{center}
\end{figure}

In Fig.~\ref{ratio}, we also present the ratio of neutron spectrum from the 4K and 20K moderator temperatures from  measurement and from  simulation. The ratio cancels some common systematic errors in the two measurements and should be more sensitive to the temperature dependence of the methane moderation physics. The intensity of cold neutrons is  increased for the lower moderator temperature as one would expect,  and the ratio  agrees quite well with the expectation from the MCNP simulation with our new scattering kernels, ``\textit{y-smeth20K}" and ``\textit{y-smeth4K}".

\section{\label{sec:conclusion} Conclusions}

We measured  the neutron energy spectrum  from the solid methane moderator of the LENS neutron source with the moderator operated in phase II at 20K and 4K.  and compared the spectra with our theoretical studies of the neutron scattering model of solid methane. We added $O_{2}$ to the solid methane to ensure spin temperature equilibrium  and we slowly cooled and thermally cycled the methane in an attempt to minimize possible cracks and holes in the solid methane. In the phase II temperatures of solid methane, 20K and 4K. The neutron energy spectrum was calculated from MCNP with scattering kernels, ``\textit{y-smeth20K}" and ``\textit{y-smeth4K}" in the geometry at LENS beam line.  The simulated neutron energy spectrum in 4K shows much colder and brighter spectrum than the 20K one. We also investigated the optimal thickness of our moderator in the temperatures with our kernels. The  MCNP simulation results were compared to the measured neutron energy spectrum. The prediction of the simulations with these newly-developed scattering kernels are in good agreement with experiment We plan to continue further spectral measurements from solid methane under different conditions in the future.
\section{\label{sec:level1} Acknowledgments}
This work was supported by the National Scientific Foundation under Grant No. DMR-0220560 and DMR-0320627.
\newpage

\bibliography{preprint2-yun}

\end{document}